\documentclass[english,prl,reprint,twocolumn,superscriptaddress,showpacs]{revtex4-1}
\usepackage[T1]{fontenc}
\usepackage[latin9]{inputenc}
\usepackage{amsmath}
\usepackage{amssymb}
\usepackage{graphicx}
\usepackage{esint}

\makeatletter
\@ifundefined{textcolor}{}
{%
 \definecolor{BLACK}{gray}{0}
 \definecolor{WHITE}{gray}{1}
 \definecolor{RED}{rgb}{1,0,0}
 \definecolor{GREEN}{rgb}{0,1,0}
 \definecolor{BLUE}{rgb}{0,0,1}
 \definecolor{CYAN}{cmyk}{1,0,0,0}
 \definecolor{MAGENTA}{cmyk}{0,1,0,0}
 \definecolor{YELLOW}{cmyk}{0,0,1,0}
}

\usepackage{siunitx}
\usepackage{mhchem}

\newcommand{\SE}{Schr\"odinger equation}
\renewcommand\[{\begin{equation}}
\renewcommand\]{\end{equation}} 

\PassOptionsToPackage{caption=false}{subfig}

\makeatother

\usepackage{babel}
\begin{document}

\title{The non-Abelian geometric phase in the diamond nitrogen-vacancy center}

\author{Mark A. Kowarsky}

\affiliation{School of Physics, The University of Melbourne, Parkville, 3010,
Australia}

\affiliation{Center for Quantum Computation and Communication Technology, School
of Physics, The University of Melbourne, Parkville, 3010, Australia}

\affiliation{Department of Physics, Stanford University, Stanford, CA 94305, USA}

\author{Lloyd C. L. Hollenberg}

\affiliation{School of Physics, The University of Melbourne, Parkville, 3010,
Australia}

\affiliation{Center for Quantum Computation and Communication Technology, School
of Physics, The University of Melbourne, Parkville, 3010, Australia}

\author{Andrew M. Martin}

\affiliation{School of Physics, The University of Melbourne, Parkville, 3010,
Australia}
\begin{abstract}
This paper introduces a theoretical framework for understanding the
accumulation of non-Abelian geometric phases in rotating nitrogen-vacancy
centers in diamond. Specifically, we consider how degenerate states
can be achieved and demonstrate that the resulting geometric phase
for multiple paths is non-Abelian. We find that the non-Abelian nature
of the phase is robust to fluctuations in the path and magnetic field.
In contrast to previous studies of the accumulation of Abelian geometric
phases for nitrogen-vacancy centers under rotation we find that the
limiting time-scale is $T_{1}$. As such a non-Abelian geometric phase
accumulation in nitrogen-vacancy centers has potential advantages
for applications as gyroscopes.
\end{abstract}
\maketitle
Non-dynamical quantum phases are at the core of our understanding
of the quantum view of the world, with historical origins going back
to the work of Aharonov and Bohm\cite{aharonov1959significance} in
1959. In 1984 Berry developed an elegant and powerful mathematical
framework\cite{berry1984quantal,simon1983holonomy} that established
the Aharonov-Bohm effect as an instance of a far more general class
of phenomena\cite{aharonov1987phase}. This theory showed that for
non-degenerate systems under an adiabatic evolution\cite{messiah1962quantum}
of the Hamiltonian, an Abelian geometric phase is acquired\cite{anandan1992geometric,leek2007observation,mottonen2008experimental}.
Systems with degenerate energy levels can possess a Berry phase with
a non-Abelian structure\cite{wilczek1984appearance}. This means that
different paths of the Hamiltonian produce a geometric phase that
in general does not commute, allowing for richer dynamics and providing
a platform to implement holonomic quantum computation\cite{zanardi1999holonomic,jones2000geometric}.
Recently an experiment using a superconducting circuit found unambiguous
evidence\cite{zee1988non,zhang2008detecting} of the non-Abelian nature
of the phase\cite{abdumalikov2013experimental}.

\begin{figure}
\includegraphics[width=0.85\columnwidth]{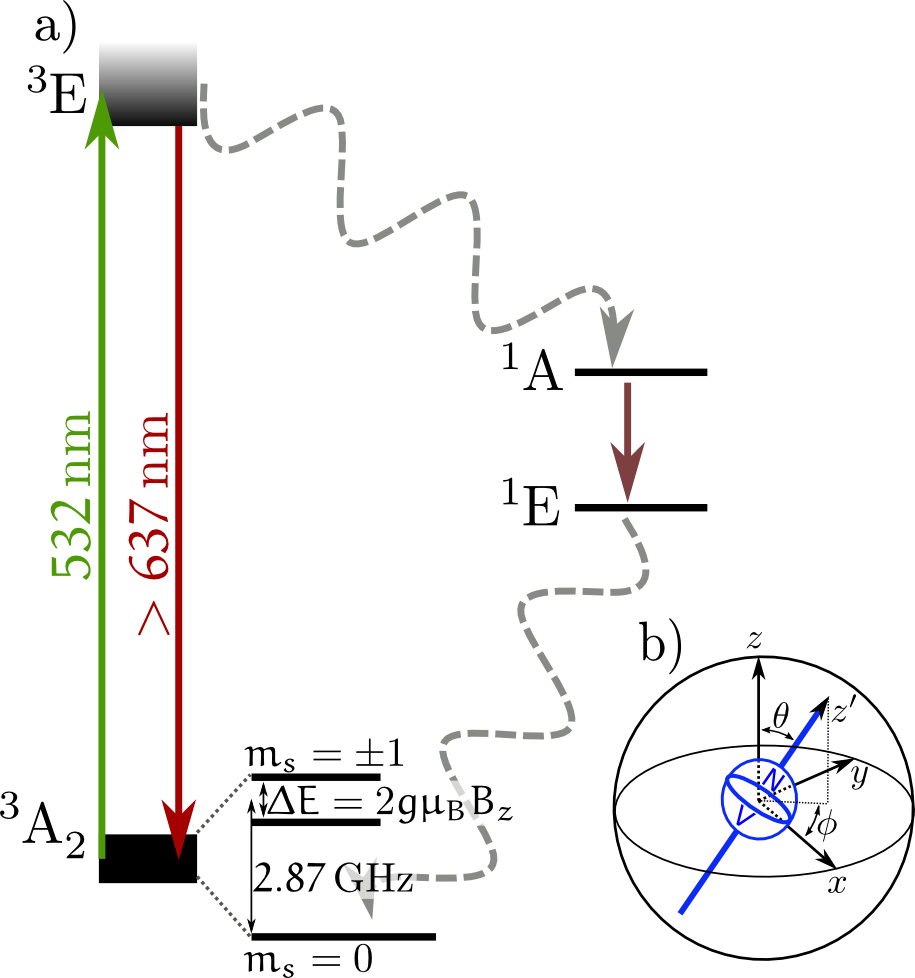}\caption{a) Energy level diagram of the NV center b) Geometry of the NV center.
Defining the microwave pulses as the $z$ direction, $z'$ is the
instantaneous direction of the NV axis, defined with respect to the
lab frame, unprimed coordinate system, by $\theta$ and $\phi$ \label{fig:introduction}}
\vspace{-0.25cm}
\end{figure}

The nitrogen-vacancy (NV) center in diamond offers a robust and accessible
single-spin system with applications in quantum communications\cite{gaebel2004stable,beveratos2002room},
quantum information\cite{gaebel2006room}, nanoscale magnetometry\cite{balasubramanian2008nanoscale,maze2008nanoscale,taylor2008high,degen2008scanning,cole2009scanning,maletinsky2012robust,steinert2013magnetic,mamin2013nanoscale},
biosensing\cite{mcguinness2011quantum,hall2010monitoring,hall2012high,pham2011magnetic,kaufmann2013detection}
and thermometry\cite{neumann2013high,toyli2013fluorescence,kucsko2013nanometre}.
The NV center (for a review see Ref \cite{doherty2013nitrogen}) is
a defect in diamond whereby a carbon is replaced by a nitrogen and
an adjacent carbon is removed. It behaves as an electronic \ce{^3A_2}
spin triplet system in the ground state, with an excited \ce{^3E}
state and a metastable \ce{^1A_1} state, see Fig. \ref{fig:introduction}a).
A laser with a wavelength shorter than the ZPL (\SI{637}{\nano\metre})
polarizes the system into the $m_{s}=0$ ground state and also allows
the spin to be read out via the fluorescence intensity. The ground
state has relatively long coherence times, even at room temperature,
with the inhomogeneous broadening time $T_{2}^{\star}$ of the order
of \si{\micro\second} and the homogeneous broadening and spin relaxation
times $T_{2}$ and $T_{1}$ of the order of \si{\milli\second}\cite{jelezko2004observation,taylor2008high,balasubramanian2009ultralong}.
Recently there has been work analysing the emergence of the Abelian
geometric phase in the NV center in rotating systems\cite{maclaurin2012measurable,ledbetter2012gyroscopes2,maclaurin2013nanoscale}
which could lead to using them as nanoscale gyroscopes. The ability
to manipulate the magnetic sub-levels with an external magnetic field
enables the possibility of degeneracy between all possible pairs of
eigenstates. As such, a single NV system provides an ideal platform
to study non-Abelian phases.

In this work, we show that the unique properties afforded to the NV
center enable the interrogation of the non-Abelian quantum phase.
It is compared with the Abelian case with limits on the angular sensitivity
derived. The chief advantage of working in the non-Abelian regime
is that the limiting coherence time is extended to $T_{1}$, whereas
in the Abelian case the measurements are limited by $T_{2}$ or $T_{2}^{*}$.
Such measurements would provide a platform to implement NV centers
as rotational sensors that are relatively insensitive to the magnetic
field noise. .

\begin{figure}
\includegraphics[width=0.6\columnwidth]{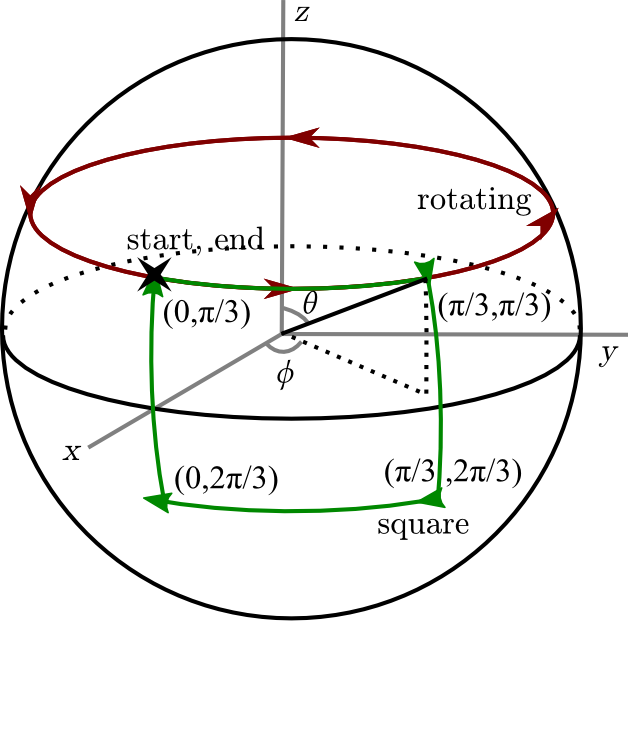}\caption{The two paths of rotation of the NV center considered in this paper.
Both start and end in the same orientation.\label{fig:paths}}
\vspace{-0.6cm}
\end{figure}

A general non-Abelian Berry phase can be understood in terms of a
Hamiltonian with $N$ degenerate eigenstates $|a(\vec{\lambda})\rangle$,
written in terms of parameters $\vec{\lambda}$ that undergo an adiabatic
evolution\cite{zee1988non}. For an initial state given by a coherent
superposition of degenerate energy eigenstates, the time evolution
operator is

\[
U=\mathcal{P}\exp\left(-\int A_{\alpha}d\lambda^{\alpha}\right),
\]
where $\mathcal{P}$ is the path ordering operator and $\alpha$ is
summed over the parameters, for example $\vec{\lambda}=(\lambda^{1},\lambda^{2})=(\theta,\phi)$.
The effective gauge potential $A_{\alpha}$, is an $N\times N$ matrix:
\begin{equation}
A_{ab\alpha}=\langle a(\vec{\lambda})|\frac{\partial}{\partial\lambda^{\alpha}}|b(\vec{\lambda})\rangle,\label{eq:effective gauge potential}
\end{equation}
where $a$ and $b$ label the degenerate eigenstates. The effect of
$U$ in general will cause a mixing between degenerate eigenstates,
and unlike in the non-degenerate $U(1)$ case, the phase cannot be
detected directly but only the trace or eigenvalues of $U$ can be
determined through a population measurement. To rigorously demonstrate
the non-Abelian nature of the acquired phase, two paths in parameter
space can be considered (for example those in Fig. \ref{fig:paths},
labelled $1$ and $2$), with equal beginning and end points. The
non-Abelian nature is seen by comparing $U=U_{2}U_{1}$ with $U^{\prime}=U_{1}U_{2}$.
In the Abelian case $U=U^{\prime}$ whereas in the non-Abelian case
in general $U\neq U^{\prime}$. WE now consider the specifics of the
NV center.

The Hamiltonian for the NV spin triplet ground state system in the
limit of a low strain diamond and with negligible hyperfine coupling
is given by
\[
H^{\prime}=DS_{z^{\prime}}^{2}+\gamma\mathbf{B}\cdot\mathbf{S},
\]
 where $z{}^{\prime}$ is the axis from the nitrogen atom to the adjacent
vacancy, the applied magnetic field is $\mathbf{B}$, spin operator
$\mathbf{S}$ and $\gamma$ is the gyromagnetic ratio of the NV center,
$D\approx$ \SI{2.87}{\giga\hertz} is the zero field splitting and
$\hbar=1$. In the lab frame, given by unprimed coordinates, the Hamiltonian
takes the form $H=RH'R^{-1}$, where $R$ is the rotation operator
$R=\exp(-i\phi S_{z})\exp(-i\theta S_{y})\exp(i\phi S_{z})$, where
$\theta$ and $\phi$ map between the $z$ and $z^{\prime}$ axes
{[}see Fig. \ref{fig:introduction}b){]}. With no applied field, the
$m_{s}=\pm1$ states are degenerate. Applying a magnetic field along
the $z^{\prime}$ axis induces a Zeeman shift of $\Delta E=\pm\gamma B_{z'}$.
A normalized Hamiltonian will be considered from here on, where $H\to H/D$
and $\epsilon\equiv\gamma B_{z}/D$ is a measure of the separation
of the energy levels. Applying a field of $\epsilon=\pm1$ makes the
$m_{s}=\mp1$ and $m_{s}=0$ states degenerate.

The effective gauge potential can be calculated for all pairs of states
using Eq. (\ref{eq:effective gauge potential}). For zero applied
field, the gauge potential ($A$ is defined as $\sum_{\alpha}A_{\alpha}d\lambda^{\alpha}$)
has the following form in the $\{|1\rangle,|-1\rangle\}$ basis
\[
A=\left(\begin{array}{cc}
-i\cos\theta d\phi & 0\\
0 & i\cos\theta d\phi
\end{array}\right).
\]
This matrix is Abelian because only entries with $d\phi$ are non-zero.
In fact, upon integration around a path, the diagonal entries are
proportional to the solid angle enclosed and the phase is identical
to the Abelian Berry phase\cite{maclaurin2012measurable,ledbetter2012gyroscopes2}.
Applying a suitable magnetic field ($\epsilon=\mp1$) along the $z^{\prime}$-axis
results in a gauge potential in the $\{|\pm1\rangle,|0\rangle\}$
basis that has a non-Abelian nature, due to the presence of both $d\phi$
and $d\theta$ terms,

\[
A=\begin{pmatrix}\mp i\cos\theta d\phi & \frac{1}{\sqrt{2}}(i\sin\theta d\phi\mp d\theta)\\
\frac{1}{\sqrt{2}}(i\sin\theta d\phi\pm d\theta) & 0
\end{pmatrix}.
\]

To unambiguously demonstrate the non-Abelian nature of the phase at
we first consider a situation when the $|0\rangle$ and $|1\rangle$
states are perfectly degenerate and the paths are exactly those as
shown in Fig. \ref{fig:paths}. To maintain the degeneracy of the
$|0\rangle$ and the $|1\rangle$ states, the crystal could be affixed
to a magnet that supplies the constant magnetic field such that $\epsilon=-1$.
The compound system could then be placed on a spinning device such
that the crystal and magnet rotate together and degeneracy is maintained.

Before the whole path is considered, the mixing effect for sub-paths
is examined by calculating the form of $U$ whilst holding one of
$\theta$ or $\phi$ constant. For $d\phi=0$ the geometric phase
in the $\{|\pm1\rangle,|0\rangle\}$ basis is 
\begin{equation}
\exp\left(-\int_{\Theta}A\right)=\begin{pmatrix}\cos(\frac{\Theta}{\sqrt{2}}) & \sin(\frac{\Theta}{\sqrt{2}})\\
-\sin(\frac{\Theta}{\sqrt{2}}) & \cos(\frac{\Theta}{\sqrt{2}})
\end{pmatrix},\label{eq:U longitude}
\end{equation}
where $\Theta=\int d\theta$ is the polar angle through which the
state is rotated. This can be understood in the following manner:
a physical rotation of the crystal through an angle of $\Theta$ induces
a rotation in the eigenspace of $-\Theta/\sqrt{2}$, independent of
$\phi$. In contrast, when $d\theta=0$ the behaviour is dependent
on $\theta$. Setting $\theta=\pi/3$ for simplicity and rotating
through an azimuthal angle $\Phi$:\begin{widetext}
\begin{equation}
\exp\left(-\int_{\Phi}A\right)=e^{i\Phi/4}\begin{pmatrix}\cos(\frac{\sqrt{7}\Phi}{4})+\frac{i}{\sqrt{7}}\sin(\frac{\sqrt{7}\Phi}{4}) & -i\sqrt{\frac{6}{7}}\sin(\frac{\sqrt{7}\Phi}{4})\\
-i\sqrt{\frac{6}{7}}\sin(\frac{\sqrt{7}\Phi}{4}) & \cos(\frac{\sqrt{7}\Phi}{4})-\frac{i}{\sqrt{7}}\sin(\frac{\sqrt{7}\Phi}{4})
\end{pmatrix}.\label{eq:U lattitude}
\end{equation}
\end{widetext}Expressing this in terms of the Pauli matrices, Eq.
(\ref{eq:U lattitude}) can be thought of as a rotation about the
axis that makes the angle $\arctan(1/\sqrt{6})$ from the negative
$x$-axis to the $z$-axis. With these two segments of the paths considered,
the total effect of the two paths shown in Fig. \ref{fig:paths} can
be evaluated. The first is a rotation around the sphere at $\theta=\pi/3$.
The phase matrix for this is given by Eq. (\ref{eq:U lattitude})
where $\Phi=2\pi$. The second path is a \emph{square} in the space
defined by $(\phi,\theta)$. Each leg of the path travels along lines
of constant latitude or longitude between the points given by ($\phi,\theta$)=\{($0,\pi/3$),
($\pi/3,\pi/3$), ($\pi/3,2\pi/3$), ($0,2\pi/3$)\}. Over each of
these sub-paths the phase accumulated has an Abelian nature, since
only one of $d\phi$ or $d\theta$ are non-zero over the length of
the path. The path integration can be done analytically, but its form
is not concise or enlightening. A numerical approximation of it is
\[
U_{\mathrm{square}}\approx\begin{pmatrix}0.91+0.23i & -0.11-0.33i\\
0.34-0.07i & 0.66+0.67i
\end{pmatrix}.
\]
Since both paths start and finish at the same point, they offer the
potential to show unambiguously the non-Abelian nature of the Berry
phase. If the system is initially placed in the $m_{s}=1$ state and
traverses the two paths in one order and population of the $m_{s}=1$
state is measured, then the experiment is repeated with the opposite
order of the paths, the final population difference between the two
paths amounts to 14.4\%. This is not the optimal contrast, but demonstrates
that for these paths chosen for analytical convenience, the non-Abelian
effect is present.

The analysis above does not deal with experimental considerations
such as decoherence, imperfect degeneracies and whether the evolution
is adiabatic. Below we demonstrate that the non-Abelian phase can
be measured even in non-ideal systems.
\begin{figure}
\includegraphics[width=0.85\columnwidth]{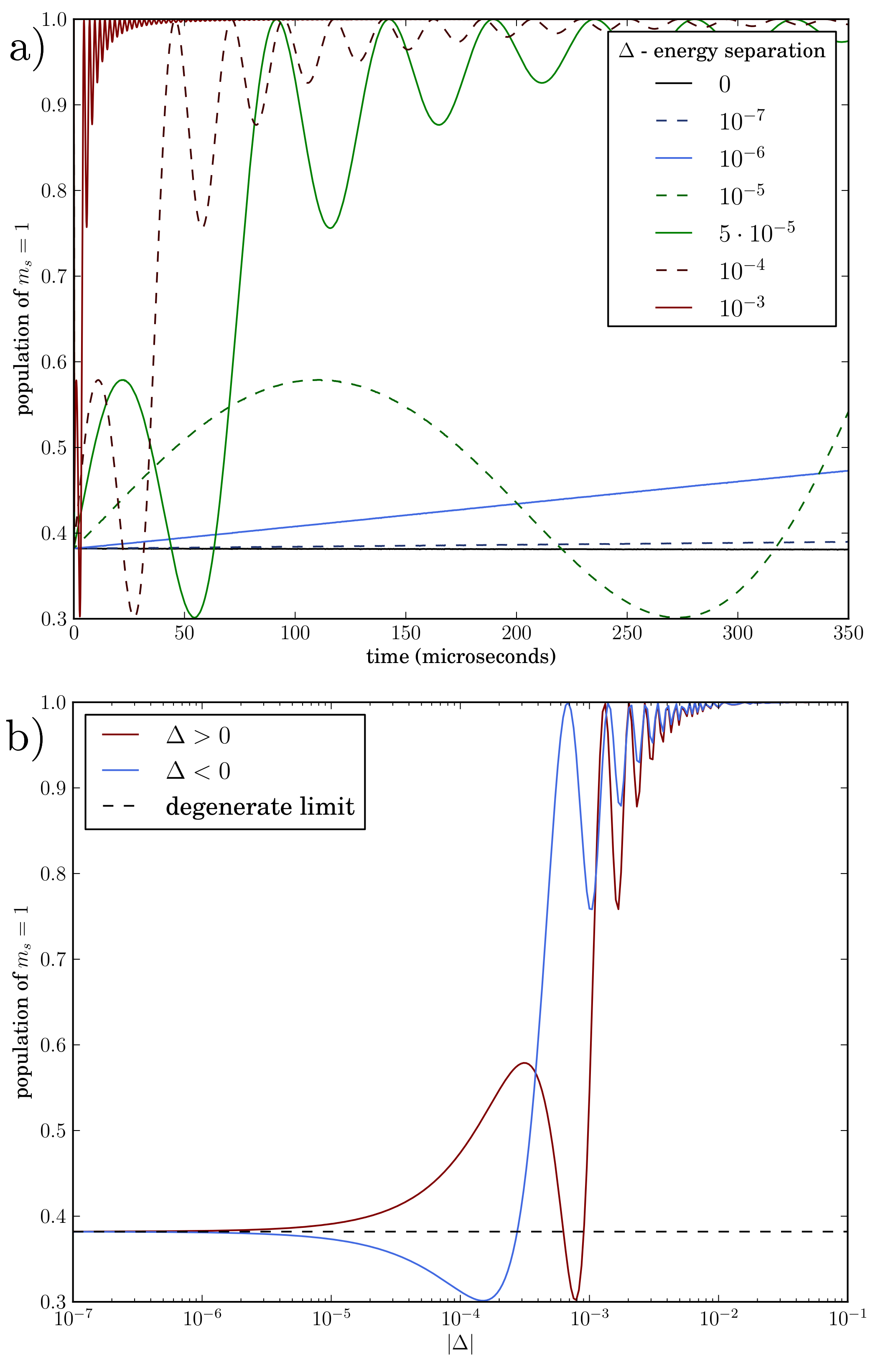}\caption{a) The population in the $m_{s}=1$ state after evolution in the rotating
path for varying degrees of degeneracy as a function of rotation time.
The asymptotic degenerate result (solid black line) corresponds exactly
with the expected adiabatic result.\label{fig:adiabatic limit} b)
The effect of degeneracy for a rotation of $\approx$\SI{3.5}{\micro\second}
on the population in $m_{s}=1$. Dashed line is the result expected
for perfect degeneracy, which is achieved for within 5\% for $|\Delta|<2\times10^{-5}$.\label{fig:nonabelianness}}
\vspace{-0.75cm}
\end{figure}

To investigate the effect of imperfect degeneracy, the \SE{ } ($\hbar=1$)
is written in terms of the the reduced time $s=t/T$, $i\frac{d}{ds}|\psi(s)\rangle=TH(s)|\psi(s)\rangle$,
where $T$ is the total time taken for the evolution. The general
solution to this equation is $|\psi(s)\rangle=\mathcal{T}\exp\left(-iT\int H(s)ds\right)|\psi(0)\rangle$,
where $\mathcal{T}$ is the time-ordering operator. Using the rotating
path defined in the previous section (see Fig. \ref{fig:paths}),
this was numerically solved for different values of $T$ and of the
energy separation $\mbox{\ensuremath{\Delta}=1+\ensuremath{\epsilon}}$
for $\epsilon\approx-1$ (for $\epsilon=-1$ the $m_{s}=0$ and $1$
states are degenerate). These calculations, for $\Phi=2\pi$ are presented
in Fig. \ref{fig:adiabatic limit}a) as a function of rotation time.
The maximum rotation time considered is \SI{350}{\micro\second} below
(above) usual values for $T_{1}$ ($T_{2}^{\star}$) of \si{\milli\second}
(\si{\micro\second})\cite{jelezko2004observation,taylor2008high,balasubramanian2009ultralong}
and also within potentially achievable \si{\kilo\hertz} range rotational
frequencies. For larger energy separations ($\Delta>10^{-4}$) the
state quickly reaches the result expected for a non-degenerate adiabatic
process. As $\Delta$ is reduced, the first ``dip'' extends for
a longer period, getting closer and closer to the result expected
for true degeneracy (solid black line). For a fixed time, the behaviour
over many orders of magnitude of degeneracy were also considered,
see in Fig. \ref{fig:nonabelianness}b). For $|\Delta|<2\times10^{-5}$
we find that the population of the $m_{s}=1$ state is within 5\%
of the degenerate value {[}dashed line in Fig. \ref{fig:nonabelianness}){]},
after a complete rotation. This limit scales with evolution time and
the zero-field splitting. This enforces the fact that perfect degeneracy
is not required, nor are extremely long periods of evolution. All
that is required is that the time of evolution is fast compared to
the near degeneracy, and slow compared to the third state. 
\begin{figure}
\includegraphics[width=0.8\columnwidth]{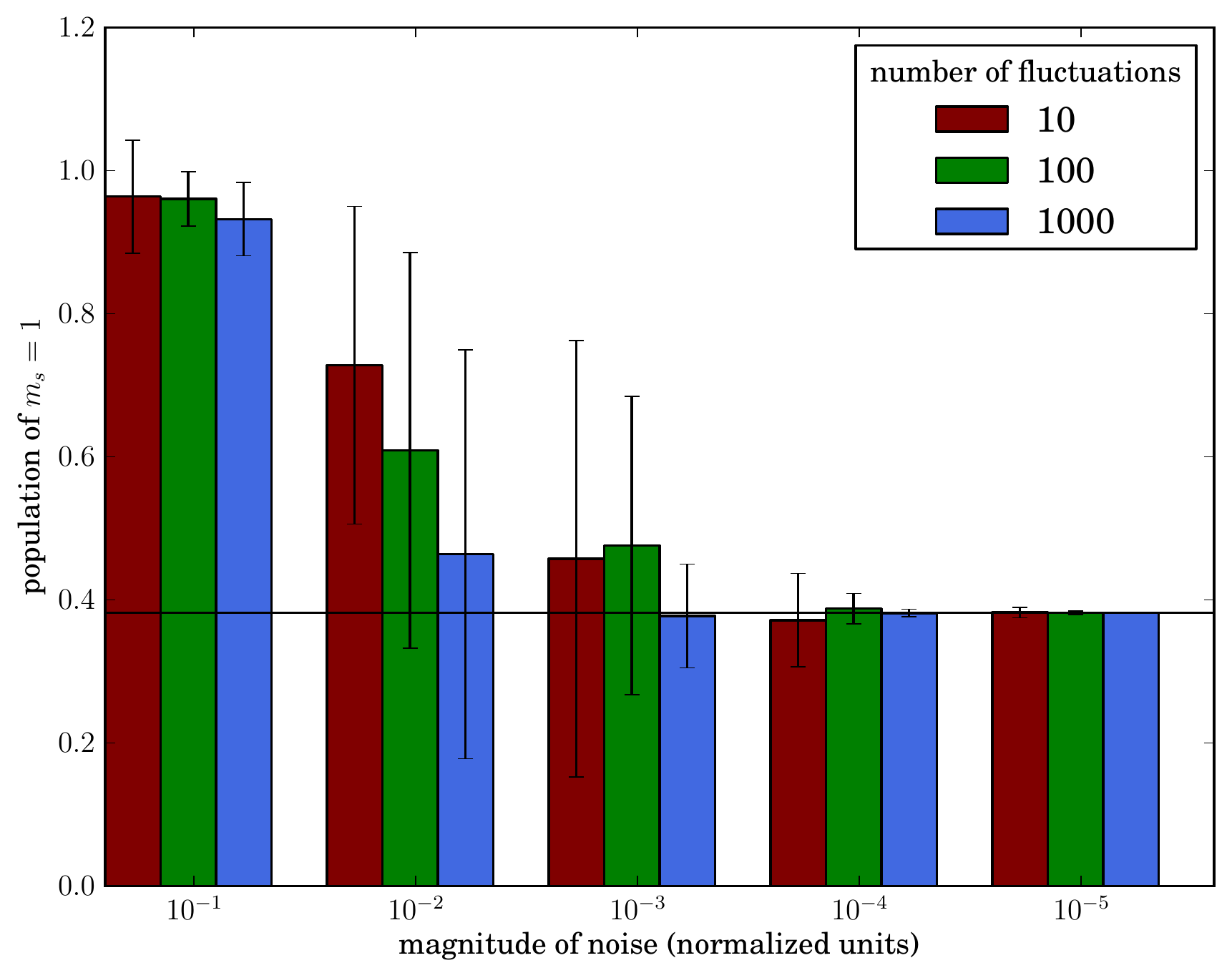} \caption{The effect that on-axis fluctuations, such as those due to $T_{2}$
processes has on the evolution (error bars are $\pm$ one standard
deviation). For levels of noise below \num{1e-4}, equivalent for
$T=10000$ (\SI{3.5}{\micro\second}) to approximately 0.1 gauss,
the effect of quickly fluctuating fields is negligible on the final
system population.\label{fig:fluctuating fields}}
\vspace{-0.75cm}
\end{figure}

Another experimental aspect that needs to be considered is the effect
that fluctuations in the magnetic field has on the evolution of the
system. An ensemble of 50 NV systems was simulated in the degenerate
limit, with Gaussian white noise fluctuations in the field along the
NV axis for a range of magnitudes and number of events over the course
of the rotating evolution. The results are summarized in Fig. \ref{fig:fluctuating fields}.
The primary effect of small stray fields is to break the degeneracy
and induce a difference in the dynamic phase between the states considered.
Since the non-Abelian experiment happens in the regime whereby the
nearly degenerate states mix, this additional $U(1)$ phase does not
influence the $U(2)$ evolution. This is markedly different to the
Abelian experiment where fluctuations in fields increase the variance
in the dynamic phase and make the geometric phase harder to recover.
This is in-line with the fact that $T_{1}$, the spin relaxation time
is the limit for measurements, not $T_{2}^{\star}$ or $T_{2}$ as
in the Abelian experimental design.

Besides investigating the effects of a non-ideal degeneracy and non-adiabatic
motion, we also considered the effects of non-ideal paths. A perfectly
known path is unobtainable experimentally so in order to determine
how errors in the path affect the measurement, perturbations away
from the expected polar angle of $\pi/3$ were simulated for the rotating
path. To remain within 5\% of the expected value, the angular divergence
required was found to be within \ang{2}. In general, unlike the Abelian
case which is very robust against classical fluctuations in the path\cite{de2003berry},
changes in the path have the potential to significantly affect the
measurement, as different paths mix the states in non-commutative
ways.

Proving the non-Abelian nature experimentally is a worthy goal, but
for a single path this approach can be applied to the NV center to
use it as a gyroscope. Consider a system where the $\theta=\pi/2$,
from Eq. (\ref{eq:U longitude}) after initialization and some rotation,
the population remaining in the state and hence the fluorescence will
be proportional to $\cos^{2}(\omega t/\sqrt{2})$, where $\omega$
is the frequency of rotation and $t$ is time. For an ensemble of
$N$ centers with a collection efficiency of $\eta$ and contrast
of $R$ between the $m_{s}=\pm1$ and $m_{s}=0$ state, the signal
is given by $F=N\eta\left(1-R\sin^{2}(\omega t/\sqrt{2})\right).$
The smallest detectable frequency is given by $\delta\omega=(dF/d\omega)^{-1}\delta F$,
where $\delta F=\sqrt{N\eta}$ is photon shot noise. For a suitable
$t$, $dF/d\omega=\sqrt{2}N\eta Rt$ and for multiple measurements
of time $\tau$ up until the limit $t=T_{1}$, the smallest frequency
can be written as
\[
\delta\omega\approx1/\alpha R\sqrt{N\eta T_{2}^{\star}\tau},
\]
where $\alpha=\sqrt{2T_{1}/T_{2}^{\star}}>1$ is the improvement factor
over the Abelian scheme ($\alpha\equiv1$) which is predicted to have
a sensitivity of \SI{5.4e-3}{\radian\per\second\per\hertz^{1/2}}\cite{ledbetter2012gyroscopes2}.
In general, $T_{1}$ is significantly longer than $T_{2}^{\star}$\cite{jelezko2004observation,taylor2008high,balasubramanian2009ultralong}
and as such it is predicted that the sensitivity can be improved by
an order of magnitude. This discussion has so far focused on the electronic
spin, but it should be possible to use the \ce{^{14}N} nuclear spin
in a similar fashion as it to is spin-1\cite{clarke1982triplet,maclaurin2012measurable,ledbetter2012gyroscopes2}.

For the NV center, the required control of the Hamiltonian is carried
out by rotating the diamond in physical space. With no applied magnetic
field, a Ramsey pulse sequence allows the Abelian phase to be detected.
Applying a magnetic field along the NV axis such that the $m_{s}=0$
state is degenerate with one of $m_{s}=\pm1$ allows the non-Abelian
phase to be detected by reading out the population from the spin-dependent
fluorescence of the center. From simulations of perturbations to the
ideal motion it was found that the non-Abelian phase is robust against
decohering effects from magnetic fields. An advantage of the non-Abelian
experiment over the Abelian experiment is that the coherence time
is increased from $T_{2}^{\star}\to T_{1}$ and thus the sensitivity
to rotations is increased. We have shown that NV centers may be used
as probes for non-Abelian geometric quantum phases, which could allow
such measurements of this phase at room temperatures. Additionally,
it offers the potential to be a more sensitive gyroscope though further
research is needed to resolve signals from multiple axes into the
3-axis rotation.
\begin{acknowledgments}
LH and MK acknowledge support from the Australian Research Council under the
Centre of Excellence scheme (project number CE110001027). MK would
like to thank D. Budker and his group for his hospitality and useful
discussions relating to extensions to this work during the later stages
of the manuscript's preparation.
\end{acknowledgments}
\bibliographystyle{apsrev4-1}
\bibliography{bibs}

\begin{thebibliography}{41}%
\makeatletter
\providecommand \@ifxundefined [1]{%
 \@ifx{#1\undefined}
}%
\providecommand \@ifnum [1]{%
 \ifnum #1\expandafter \@firstoftwo
 \else \expandafter \@secondoftwo
 \fi
}%
\providecommand \@ifx [1]{%
 \ifx #1\expandafter \@firstoftwo
 \else \expandafter \@secondoftwo
 \fi
}%
\providecommand \natexlab [1]{#1}%
\providecommand \enquote  [1]{``#1''}%
\providecommand \bibnamefont  [1]{#1}%
\providecommand \bibfnamefont [1]{#1}%
\providecommand \citenamefont [1]{#1}%
\providecommand \href@noop [0]{\@secondoftwo}%
\providecommand \href [0]{\begingroup \@sanitize@url \@href}%
\providecommand \@href[1]{\@@startlink{#1}\@@href}%
\providecommand \@@href[1]{\endgroup#1\@@endlink}%
\providecommand \@sanitize@url [0]{\catcode `\\12\catcode `\$12\catcode
  `\&12\catcode `\#12\catcode `\^12\catcode `\_12\catcode `\%12\relax}%
\providecommand \@@startlink[1]{}%
\providecommand \@@endlink[0]{}%
\providecommand \url  [0]{\begingroup\@sanitize@url \@url }%
\providecommand \@url [1]{\endgroup\@href {#1}{\urlprefix }}%
\providecommand \urlprefix  [0]{URL }%
\providecommand \Eprint [0]{\href }%
\providecommand \doibase [0]{http://dx.doi.org/}%
\providecommand \selectlanguage [0]{\@gobble}%
\providecommand \bibinfo  [0]{\@secondoftwo}%
\providecommand \bibfield  [0]{\@secondoftwo}%
\providecommand \translation [1]{[#1]}%
\providecommand \BibitemOpen [0]{}%
\providecommand \bibitemStop [0]{}%
\providecommand \bibitemNoStop [0]{.\EOS\space}%
\providecommand \EOS [0]{\spacefactor3000\relax}%
\providecommand \BibitemShut  [1]{\csname bibitem#1\endcsname}%
\let\auto@bib@innerbib\@empty
\bibitem [{\citenamefont {Aharonov}\ and\ \citenamefont
  {Bohm}(1959)}]{aharonov1959significance}%
  \BibitemOpen
  \bibfield  {author} {\bibinfo {author} {\bibfnamefont {Y.}~\bibnamefont
  {Aharonov}}\ and\ \bibinfo {author} {\bibfnamefont {D.}~\bibnamefont
  {Bohm}},\ }\href@noop {} {\bibfield  {journal} {\bibinfo  {journal} {Physical
  Review}\ }\textbf {\bibinfo {volume} {115}},\ \bibinfo {pages} {485}
  (\bibinfo {year} {1959})}\BibitemShut {NoStop}%
\bibitem [{\citenamefont {Berry}(1984)}]{berry1984quantal}%
  \BibitemOpen
  \bibfield  {author} {\bibinfo {author} {\bibfnamefont {M.}~\bibnamefont
  {Berry}},\ }\href@noop {} {\bibfield  {journal} {\bibinfo  {journal}
  {Proceedings of the Royal Society of London. A. Mathematical and Physical
  Sciences}\ }\textbf {\bibinfo {volume} {392}},\ \bibinfo {pages} {45}
  (\bibinfo {year} {1984})}\BibitemShut {NoStop}%
\bibitem [{\citenamefont {Simon}(1983)}]{simon1983holonomy}%
  \BibitemOpen
  \bibfield  {author} {\bibinfo {author} {\bibfnamefont {B.}~\bibnamefont
  {Simon}},\ }\href@noop {} {\bibfield  {journal} {\bibinfo  {journal}
  {Physical Review Letters}\ }\textbf {\bibinfo {volume} {51}},\ \bibinfo
  {pages} {2167} (\bibinfo {year} {1983})}\BibitemShut {NoStop}%
\bibitem [{\citenamefont {Aharonov}\ and\ \citenamefont
  {Anandan}(1987)}]{aharonov1987phase}%
  \BibitemOpen
  \bibfield  {author} {\bibinfo {author} {\bibfnamefont {Y.}~\bibnamefont
  {Aharonov}}\ and\ \bibinfo {author} {\bibfnamefont {J.}~\bibnamefont
  {Anandan}},\ }\href@noop {} {\bibfield  {journal} {\bibinfo  {journal}
  {Physical Review Letters}\ }\textbf {\bibinfo {volume} {58}},\ \bibinfo
  {pages} {1593} (\bibinfo {year} {1987})}\BibitemShut {NoStop}%
\bibitem [{\citenamefont {Messiah}(1961)}]{messiah1962quantum}%
  \BibitemOpen
  \bibfield  {author} {\bibinfo {author} {\bibfnamefont {A.}~\bibnamefont
  {Messiah}},\ }\href@noop {} {\emph {\bibinfo {title} {{Quantum Mechanics
  Volume II}}}}\ (\bibinfo  {publisher} {Elsevier Science B.V.},\ \bibinfo
  {year} {1961})\BibitemShut {NoStop}%
\bibitem [{\citenamefont {Anandan}(1992)}]{anandan1992geometric}%
  \BibitemOpen
  \bibfield  {author} {\bibinfo {author} {\bibfnamefont {J.}~\bibnamefont
  {Anandan}},\ }\href@noop {} {\bibfield  {journal} {\bibinfo  {journal}
  {Nature}\ }\textbf {\bibinfo {volume} {360}},\ \bibinfo {pages} {307}
  (\bibinfo {year} {1992})}\BibitemShut {NoStop}%
\bibitem [{\citenamefont {Leek}\ \emph {et~al.}(2007)\citenamefont {Leek},
  \citenamefont {Fink}, \citenamefont {Blais}, \citenamefont {Bianchetti},
  \citenamefont {G{\"o}ppl}, \citenamefont {Gambetta}, \citenamefont
  {Schuster}, \citenamefont {Frunzio}, \citenamefont {Schoelkopf},\ and\
  \citenamefont {Wallraff}}]{leek2007observation}%
  \BibitemOpen
  \bibfield  {author} {\bibinfo {author} {\bibfnamefont {P.}~\bibnamefont
  {Leek}}, \bibinfo {author} {\bibfnamefont {J.}~\bibnamefont {Fink}}, \bibinfo
  {author} {\bibfnamefont {A.}~\bibnamefont {Blais}}, \bibinfo {author}
  {\bibfnamefont {R.}~\bibnamefont {Bianchetti}}, \bibinfo {author}
  {\bibfnamefont {M.}~\bibnamefont {G{\"o}ppl}}, \bibinfo {author}
  {\bibfnamefont {J.}~\bibnamefont {Gambetta}}, \bibinfo {author}
  {\bibfnamefont {D.}~\bibnamefont {Schuster}}, \bibinfo {author}
  {\bibfnamefont {L.}~\bibnamefont {Frunzio}}, \bibinfo {author} {\bibfnamefont
  {R.}~\bibnamefont {Schoelkopf}}, \ and\ \bibinfo {author} {\bibfnamefont
  {A.}~\bibnamefont {Wallraff}},\ }\href@noop {} {\bibfield  {journal}
  {\bibinfo  {journal} {Science}\ }\textbf {\bibinfo {volume} {318}},\ \bibinfo
  {pages} {1889} (\bibinfo {year} {2007})}\BibitemShut {NoStop}%
\bibitem [{\citenamefont {M{\"o}tt{\"o}nen}\ \emph {et~al.}(2008)\citenamefont
  {M{\"o}tt{\"o}nen}, \citenamefont {Vartiainen},\ and\ \citenamefont
  {Pekola}}]{mottonen2008experimental}%
  \BibitemOpen
  \bibfield  {author} {\bibinfo {author} {\bibfnamefont {M.}~\bibnamefont
  {M{\"o}tt{\"o}nen}}, \bibinfo {author} {\bibfnamefont {J.}~\bibnamefont
  {Vartiainen}}, \ and\ \bibinfo {author} {\bibfnamefont {J.}~\bibnamefont
  {Pekola}},\ }\href@noop {} {\bibfield  {journal} {\bibinfo  {journal}
  {Physical Review Letters}\ }\textbf {\bibinfo {volume} {100}},\ \bibinfo
  {pages} {177201} (\bibinfo {year} {2008})}\BibitemShut {NoStop}%
\bibitem [{\citenamefont {Wilczek}\ and\ \citenamefont
  {Zee}(1984)}]{wilczek1984appearance}%
  \BibitemOpen
  \bibfield  {author} {\bibinfo {author} {\bibfnamefont {F.}~\bibnamefont
  {Wilczek}}\ and\ \bibinfo {author} {\bibfnamefont {A.}~\bibnamefont {Zee}},\
  }\href@noop {} {\bibfield  {journal} {\bibinfo  {journal} {Physical Review
  Letters}\ }\textbf {\bibinfo {volume} {52}},\ \bibinfo {pages} {2111}
  (\bibinfo {year} {1984})}\BibitemShut {NoStop}%
\bibitem [{\citenamefont {Zanardi}\ and\ \citenamefont
  {Rasetti}(1999)}]{zanardi1999holonomic}%
  \BibitemOpen
  \bibfield  {author} {\bibinfo {author} {\bibfnamefont {P.}~\bibnamefont
  {Zanardi}}\ and\ \bibinfo {author} {\bibfnamefont {M.}~\bibnamefont
  {Rasetti}},\ }\href@noop {} {\bibfield  {journal} {\bibinfo  {journal}
  {Physics Letters A}\ }\textbf {\bibinfo {volume} {264}},\ \bibinfo {pages}
  {94} (\bibinfo {year} {1999})}\BibitemShut {NoStop}%
\bibitem [{\citenamefont {Jones}\ \emph {et~al.}(2000)\citenamefont {Jones},
  \citenamefont {Vedral}, \citenamefont {Ekert}, \citenamefont {Castagnoli}
  \emph {et~al.}}]{jones2000geometric}%
  \BibitemOpen
  \bibfield  {author} {\bibinfo {author} {\bibfnamefont {J.}~\bibnamefont
  {Jones}}, \bibinfo {author} {\bibfnamefont {V.}~\bibnamefont {Vedral}},
  \bibinfo {author} {\bibfnamefont {A.}~\bibnamefont {Ekert}}, \bibinfo
  {author} {\bibfnamefont {G.}~\bibnamefont {Castagnoli}},  \emph {et~al.},\
  }\href@noop {} {\bibfield  {journal} {\bibinfo  {journal} {Nature}\ }\textbf
  {\bibinfo {volume} {403}},\ \bibinfo {pages} {869} (\bibinfo {year}
  {2000})}\BibitemShut {NoStop}%
\bibitem [{\citenamefont {Zee}(1988)}]{zee1988non}%
  \BibitemOpen
  \bibfield  {author} {\bibinfo {author} {\bibfnamefont {A.}~\bibnamefont
  {Zee}},\ }\href@noop {} {\bibfield  {journal} {\bibinfo  {journal} {Physical
  Review A}\ }\textbf {\bibinfo {volume} {38}},\ \bibinfo {pages} {1} (\bibinfo
  {year} {1988})}\BibitemShut {NoStop}%
\bibitem [{\citenamefont {Zhang}\ \emph {et~al.}(2008)\citenamefont {Zhang},
  \citenamefont {Wang}, \citenamefont {Hu}, \citenamefont {Zhang},\ and\
  \citenamefont {Zhu}}]{zhang2008detecting}%
  \BibitemOpen
  \bibfield  {author} {\bibinfo {author} {\bibfnamefont {X.}~\bibnamefont
  {Zhang}}, \bibinfo {author} {\bibfnamefont {Z.}~\bibnamefont {Wang}},
  \bibinfo {author} {\bibfnamefont {L.}~\bibnamefont {Hu}}, \bibinfo {author}
  {\bibfnamefont {Z.}~\bibnamefont {Zhang}}, \ and\ \bibinfo {author}
  {\bibfnamefont {S.}~\bibnamefont {Zhu}},\ }\href@noop {} {\bibfield
  {journal} {\bibinfo  {journal} {New Journal of Physics}\ }\textbf {\bibinfo
  {volume} {10}},\ \bibinfo {pages} {043031} (\bibinfo {year}
  {2008})}\BibitemShut {NoStop}%
\bibitem [{\citenamefont {Abdumalikov}\ \emph {et~al.}(2013)\citenamefont
  {Abdumalikov}, \citenamefont {Fink}, \citenamefont {Juliusson}, \citenamefont
  {Pechal}, \citenamefont {Berger}, \citenamefont {Wallraff},\ and\
  \citenamefont {Filipp}}]{abdumalikov2013experimental}%
  \BibitemOpen
  \bibfield  {author} {\bibinfo {author} {\bibfnamefont {A.}~\bibnamefont
  {Abdumalikov}}, \bibinfo {author} {\bibfnamefont {J.}~\bibnamefont {Fink}},
  \bibinfo {author} {\bibfnamefont {K.}~\bibnamefont {Juliusson}}, \bibinfo
  {author} {\bibfnamefont {M.}~\bibnamefont {Pechal}}, \bibinfo {author}
  {\bibfnamefont {S.}~\bibnamefont {Berger}}, \bibinfo {author} {\bibfnamefont
  {A.}~\bibnamefont {Wallraff}}, \ and\ \bibinfo {author} {\bibfnamefont
  {S.}~\bibnamefont {Filipp}},\ }\href@noop {} {\bibfield  {journal} {\bibinfo
  {journal} {Nature}\ } (\bibinfo {year} {2013})}\BibitemShut {NoStop}%
\bibitem [{\citenamefont {Gaebel}\ \emph {et~al.}(2004)\citenamefont {Gaebel},
  \citenamefont {Popa}, \citenamefont {Gruber}, \citenamefont {Domhan},
  \citenamefont {Jelezko},\ and\ \citenamefont {Wrachtrup}}]{gaebel2004stable}%
  \BibitemOpen
  \bibfield  {author} {\bibinfo {author} {\bibfnamefont {T.}~\bibnamefont
  {Gaebel}}, \bibinfo {author} {\bibfnamefont {I.}~\bibnamefont {Popa}},
  \bibinfo {author} {\bibfnamefont {A.}~\bibnamefont {Gruber}}, \bibinfo
  {author} {\bibfnamefont {M.}~\bibnamefont {Domhan}}, \bibinfo {author}
  {\bibfnamefont {F.}~\bibnamefont {Jelezko}}, \ and\ \bibinfo {author}
  {\bibfnamefont {J.}~\bibnamefont {Wrachtrup}},\ }\href@noop {} {\bibfield
  {journal} {\bibinfo  {journal} {New Journal of Physics}\ }\textbf {\bibinfo
  {volume} {6}},\ \bibinfo {pages} {98} (\bibinfo {year} {2004})}\BibitemShut
  {NoStop}%
\bibitem [{\citenamefont {Beveratos}\ \emph {et~al.}(2002)\citenamefont
  {Beveratos}, \citenamefont {K{\"u}hn}, \citenamefont {Brouri}, \citenamefont
  {Gacoin}, \citenamefont {Poizat},\ and\ \citenamefont
  {Grangier}}]{beveratos2002room}%
  \BibitemOpen
  \bibfield  {author} {\bibinfo {author} {\bibfnamefont {A.}~\bibnamefont
  {Beveratos}}, \bibinfo {author} {\bibfnamefont {S.}~\bibnamefont {K{\"u}hn}},
  \bibinfo {author} {\bibfnamefont {R.}~\bibnamefont {Brouri}}, \bibinfo
  {author} {\bibfnamefont {T.}~\bibnamefont {Gacoin}}, \bibinfo {author}
  {\bibfnamefont {J.-P.}\ \bibnamefont {Poizat}}, \ and\ \bibinfo {author}
  {\bibfnamefont {P.}~\bibnamefont {Grangier}},\ }\href@noop {} {\bibfield
  {journal} {\bibinfo  {journal} {The European Physical Journal D-Atomic,
  Molecular, Optical and Plasma Physics}\ }\textbf {\bibinfo {volume} {18}},\
  \bibinfo {pages} {191} (\bibinfo {year} {2002})}\BibitemShut {NoStop}%
\bibitem [{\citenamefont {Gaebel}\ \emph {et~al.}(2006)\citenamefont {Gaebel},
  \citenamefont {Domhan}, \citenamefont {Popa}, \citenamefont {Wittmann},
  \citenamefont {Neumann}, \citenamefont {Jelezko}, \citenamefont {Rabeau},
  \citenamefont {Stavrias}, \citenamefont {Greentree}, \citenamefont {Prawer}
  \emph {et~al.}}]{gaebel2006room}%
  \BibitemOpen
  \bibfield  {author} {\bibinfo {author} {\bibfnamefont {T.}~\bibnamefont
  {Gaebel}}, \bibinfo {author} {\bibfnamefont {M.}~\bibnamefont {Domhan}},
  \bibinfo {author} {\bibfnamefont {I.}~\bibnamefont {Popa}}, \bibinfo {author}
  {\bibfnamefont {C.}~\bibnamefont {Wittmann}}, \bibinfo {author}
  {\bibfnamefont {P.}~\bibnamefont {Neumann}}, \bibinfo {author} {\bibfnamefont
  {F.}~\bibnamefont {Jelezko}}, \bibinfo {author} {\bibfnamefont
  {J.}~\bibnamefont {Rabeau}}, \bibinfo {author} {\bibfnamefont
  {N.}~\bibnamefont {Stavrias}}, \bibinfo {author} {\bibfnamefont
  {A.}~\bibnamefont {Greentree}}, \bibinfo {author} {\bibfnamefont
  {S.}~\bibnamefont {Prawer}},  \emph {et~al.},\ }\href@noop {} {\bibfield
  {journal} {\bibinfo  {journal} {Nature Physics}\ }\textbf {\bibinfo {volume}
  {2}},\ \bibinfo {pages} {408} (\bibinfo {year} {2006})}\BibitemShut {NoStop}%
\bibitem [{\citenamefont {Balasubramanian}\ \emph {et~al.}(2008)\citenamefont
  {Balasubramanian}, \citenamefont {Chan}, \citenamefont {Kolesov},
  \citenamefont {Al-Hmoud}, \citenamefont {Tisler}, \citenamefont {Shin},
  \citenamefont {Kim}, \citenamefont {Wojcik}, \citenamefont {Hemmer},
  \citenamefont {Krueger} \emph {et~al.}}]{balasubramanian2008nanoscale}%
  \BibitemOpen
  \bibfield  {author} {\bibinfo {author} {\bibfnamefont {G.}~\bibnamefont
  {Balasubramanian}}, \bibinfo {author} {\bibfnamefont {I.}~\bibnamefont
  {Chan}}, \bibinfo {author} {\bibfnamefont {R.}~\bibnamefont {Kolesov}},
  \bibinfo {author} {\bibfnamefont {M.}~\bibnamefont {Al-Hmoud}}, \bibinfo
  {author} {\bibfnamefont {J.}~\bibnamefont {Tisler}}, \bibinfo {author}
  {\bibfnamefont {C.}~\bibnamefont {Shin}}, \bibinfo {author} {\bibfnamefont
  {C.}~\bibnamefont {Kim}}, \bibinfo {author} {\bibfnamefont {A.}~\bibnamefont
  {Wojcik}}, \bibinfo {author} {\bibfnamefont {P.}~\bibnamefont {Hemmer}},
  \bibinfo {author} {\bibfnamefont {A.}~\bibnamefont {Krueger}},  \emph
  {et~al.},\ }\href@noop {} {\bibfield  {journal} {\bibinfo  {journal}
  {Nature}\ }\textbf {\bibinfo {volume} {455}},\ \bibinfo {pages} {648}
  (\bibinfo {year} {2008})}\BibitemShut {NoStop}%
\bibitem [{\citenamefont {Maze}\ \emph {et~al.}(2008)\citenamefont {Maze},
  \citenamefont {Stanwix}, \citenamefont {Hodges}, \citenamefont {Hong},
  \citenamefont {Taylor}, \citenamefont {Cappellaro}, \citenamefont {Jiang},
  \citenamefont {Dutt}, \citenamefont {Togan}, \citenamefont {Zibrov} \emph
  {et~al.}}]{maze2008nanoscale}%
  \BibitemOpen
  \bibfield  {author} {\bibinfo {author} {\bibfnamefont {J.}~\bibnamefont
  {Maze}}, \bibinfo {author} {\bibfnamefont {P.}~\bibnamefont {Stanwix}},
  \bibinfo {author} {\bibfnamefont {J.}~\bibnamefont {Hodges}}, \bibinfo
  {author} {\bibfnamefont {S.}~\bibnamefont {Hong}}, \bibinfo {author}
  {\bibfnamefont {J.}~\bibnamefont {Taylor}}, \bibinfo {author} {\bibfnamefont
  {P.}~\bibnamefont {Cappellaro}}, \bibinfo {author} {\bibfnamefont
  {L.}~\bibnamefont {Jiang}}, \bibinfo {author} {\bibfnamefont
  {M.}~\bibnamefont {Dutt}}, \bibinfo {author} {\bibfnamefont {E.}~\bibnamefont
  {Togan}}, \bibinfo {author} {\bibfnamefont {A.}~\bibnamefont {Zibrov}},
  \emph {et~al.},\ }\href@noop {} {\bibfield  {journal} {\bibinfo  {journal}
  {Nature}\ }\textbf {\bibinfo {volume} {455}},\ \bibinfo {pages} {644}
  (\bibinfo {year} {2008})}\BibitemShut {NoStop}%
\bibitem [{\citenamefont {Taylor}\ \emph {et~al.}(2008)\citenamefont {Taylor},
  \citenamefont {Cappellaro}, \citenamefont {Childress}, \citenamefont {Jiang},
  \citenamefont {Budker}, \citenamefont {Hemmer}, \citenamefont {Yacoby},
  \citenamefont {Walsworth},\ and\ \citenamefont {Lukin}}]{taylor2008high}%
  \BibitemOpen
  \bibfield  {author} {\bibinfo {author} {\bibfnamefont {J.}~\bibnamefont
  {Taylor}}, \bibinfo {author} {\bibfnamefont {P.}~\bibnamefont {Cappellaro}},
  \bibinfo {author} {\bibfnamefont {L.}~\bibnamefont {Childress}}, \bibinfo
  {author} {\bibfnamefont {L.}~\bibnamefont {Jiang}}, \bibinfo {author}
  {\bibfnamefont {D.}~\bibnamefont {Budker}}, \bibinfo {author} {\bibfnamefont
  {P.}~\bibnamefont {Hemmer}}, \bibinfo {author} {\bibfnamefont
  {A.}~\bibnamefont {Yacoby}}, \bibinfo {author} {\bibfnamefont
  {R.}~\bibnamefont {Walsworth}}, \ and\ \bibinfo {author} {\bibfnamefont
  {M.}~\bibnamefont {Lukin}},\ }\href@noop {} {\bibfield  {journal} {\bibinfo
  {journal} {Nature Physics}\ }\textbf {\bibinfo {volume} {4}},\ \bibinfo
  {pages} {810} (\bibinfo {year} {2008})}\BibitemShut {NoStop}%
\bibitem [{\citenamefont {Degen}(2008)}]{degen2008scanning}%
  \BibitemOpen
  \bibfield  {author} {\bibinfo {author} {\bibfnamefont {C.}~\bibnamefont
  {Degen}},\ }\href@noop {} {\bibfield  {journal} {\bibinfo  {journal} {Applied
  Physics Letters}\ }\textbf {\bibinfo {volume} {92}},\ \bibinfo {pages}
  {243111} (\bibinfo {year} {2008})}\BibitemShut {NoStop}%
\bibitem [{\citenamefont {Cole}\ and\ \citenamefont
  {Hollenberg}(2009)}]{cole2009scanning}%
  \BibitemOpen
  \bibfield  {author} {\bibinfo {author} {\bibfnamefont {J.}~\bibnamefont
  {Cole}}\ and\ \bibinfo {author} {\bibfnamefont {L.}~\bibnamefont
  {Hollenberg}},\ }\href@noop {} {\bibfield  {journal} {\bibinfo  {journal}
  {Nanotechnology}\ }\textbf {\bibinfo {volume} {20}},\ \bibinfo {pages}
  {495401} (\bibinfo {year} {2009})}\BibitemShut {NoStop}%
\bibitem [{\citenamefont {Maletinsky}\ \emph {et~al.}(2012)\citenamefont
  {Maletinsky}, \citenamefont {Hong}, \citenamefont {Grinolds}, \citenamefont
  {Hausmann}, \citenamefont {Lukin}, \citenamefont {Walsworth}, \citenamefont
  {Loncar},\ and\ \citenamefont {Yacoby}}]{maletinsky2012robust}%
  \BibitemOpen
  \bibfield  {author} {\bibinfo {author} {\bibfnamefont {P.}~\bibnamefont
  {Maletinsky}}, \bibinfo {author} {\bibfnamefont {S.}~\bibnamefont {Hong}},
  \bibinfo {author} {\bibfnamefont {M.~S.}\ \bibnamefont {Grinolds}}, \bibinfo
  {author} {\bibfnamefont {B.}~\bibnamefont {Hausmann}}, \bibinfo {author}
  {\bibfnamefont {M.~D.}\ \bibnamefont {Lukin}}, \bibinfo {author}
  {\bibfnamefont {R.~L.}\ \bibnamefont {Walsworth}}, \bibinfo {author}
  {\bibfnamefont {M.}~\bibnamefont {Loncar}}, \ and\ \bibinfo {author}
  {\bibfnamefont {A.}~\bibnamefont {Yacoby}},\ }\href@noop {} {\bibfield
  {journal} {\bibinfo  {journal} {Nature Nanotechnology}\ }\textbf {\bibinfo
  {volume} {7}},\ \bibinfo {pages} {320} (\bibinfo {year} {2012})}\BibitemShut
  {NoStop}%
\bibitem [{\citenamefont {Steinert}\ \emph {et~al.}(2013)\citenamefont
  {Steinert}, \citenamefont {Ziem}, \citenamefont {Hall}, \citenamefont
  {Zappe}, \citenamefont {Schweikert}, \citenamefont {G{\"o}tz}, \citenamefont
  {Aird}, \citenamefont {Balasubramanian}, \citenamefont {Hollenberg},\ and\
  \citenamefont {Wrachtrup}}]{steinert2013magnetic}%
  \BibitemOpen
  \bibfield  {author} {\bibinfo {author} {\bibfnamefont {S.}~\bibnamefont
  {Steinert}}, \bibinfo {author} {\bibfnamefont {F.}~\bibnamefont {Ziem}},
  \bibinfo {author} {\bibfnamefont {L.}~\bibnamefont {Hall}}, \bibinfo {author}
  {\bibfnamefont {A.}~\bibnamefont {Zappe}}, \bibinfo {author} {\bibfnamefont
  {M.}~\bibnamefont {Schweikert}}, \bibinfo {author} {\bibfnamefont
  {N.}~\bibnamefont {G{\"o}tz}}, \bibinfo {author} {\bibfnamefont
  {A.}~\bibnamefont {Aird}}, \bibinfo {author} {\bibfnamefont {G.}~\bibnamefont
  {Balasubramanian}}, \bibinfo {author} {\bibfnamefont {L.}~\bibnamefont
  {Hollenberg}}, \ and\ \bibinfo {author} {\bibfnamefont {J.}~\bibnamefont
  {Wrachtrup}},\ }\href@noop {} {\bibfield  {journal} {\bibinfo  {journal}
  {Nature communications}\ }\textbf {\bibinfo {volume} {4}},\ \bibinfo {pages}
  {1607} (\bibinfo {year} {2013})}\BibitemShut {NoStop}%
\bibitem [{\citenamefont {Mamin}\ \emph {et~al.}(2013)\citenamefont {Mamin},
  \citenamefont {Kim}, \citenamefont {Sherwood}, \citenamefont {Rettner},
  \citenamefont {Ohno}, \citenamefont {Awschalom},\ and\ \citenamefont
  {Rugar}}]{mamin2013nanoscale}%
  \BibitemOpen
  \bibfield  {author} {\bibinfo {author} {\bibfnamefont {H.}~\bibnamefont
  {Mamin}}, \bibinfo {author} {\bibfnamefont {M.}~\bibnamefont {Kim}}, \bibinfo
  {author} {\bibfnamefont {M.}~\bibnamefont {Sherwood}}, \bibinfo {author}
  {\bibfnamefont {C.}~\bibnamefont {Rettner}}, \bibinfo {author} {\bibfnamefont
  {K.}~\bibnamefont {Ohno}}, \bibinfo {author} {\bibfnamefont {D.}~\bibnamefont
  {Awschalom}}, \ and\ \bibinfo {author} {\bibfnamefont {D.}~\bibnamefont
  {Rugar}},\ }\href@noop {} {\bibfield  {journal} {\bibinfo  {journal}
  {Science}\ }\textbf {\bibinfo {volume} {339}},\ \bibinfo {pages} {557}
  (\bibinfo {year} {2013})}\BibitemShut {NoStop}%
\bibitem [{\citenamefont {McGuinness}\ \emph {et~al.}(2011)\citenamefont
  {McGuinness}, \citenamefont {Yan}, \citenamefont {Stacey}, \citenamefont
  {Simpson}, \citenamefont {Hall}, \citenamefont {Maclaurin}, \citenamefont
  {Prawer}, \citenamefont {Mulvaney}, \citenamefont {Wrachtrup}, \citenamefont
  {Caruso} \emph {et~al.}}]{mcguinness2011quantum}%
  \BibitemOpen
  \bibfield  {author} {\bibinfo {author} {\bibfnamefont {L.}~\bibnamefont
  {McGuinness}}, \bibinfo {author} {\bibfnamefont {Y.}~\bibnamefont {Yan}},
  \bibinfo {author} {\bibfnamefont {A.}~\bibnamefont {Stacey}}, \bibinfo
  {author} {\bibfnamefont {D.}~\bibnamefont {Simpson}}, \bibinfo {author}
  {\bibfnamefont {L.}~\bibnamefont {Hall}}, \bibinfo {author} {\bibfnamefont
  {D.}~\bibnamefont {Maclaurin}}, \bibinfo {author} {\bibfnamefont
  {S.}~\bibnamefont {Prawer}}, \bibinfo {author} {\bibfnamefont
  {P.}~\bibnamefont {Mulvaney}}, \bibinfo {author} {\bibfnamefont
  {J.}~\bibnamefont {Wrachtrup}}, \bibinfo {author} {\bibfnamefont
  {F.}~\bibnamefont {Caruso}},  \emph {et~al.},\ }\href@noop {} {\bibfield
  {journal} {\bibinfo  {journal} {Nature Nanotechnology}\ }\textbf {\bibinfo
  {volume} {6}},\ \bibinfo {pages} {358} (\bibinfo {year} {2011})}\BibitemShut
  {NoStop}%
\bibitem [{\citenamefont {Hall}\ \emph {et~al.}(2010)\citenamefont {Hall},
  \citenamefont {Hill}, \citenamefont {Cole}, \citenamefont {Städler},
  \citenamefont {Caruso}, \citenamefont {Mulvaney}, \citenamefont {Wrachtrup},\
  and\ \citenamefont {Hollenberg}}]{hall2010monitoring}%
  \BibitemOpen
  \bibfield  {author} {\bibinfo {author} {\bibfnamefont {L.}~\bibnamefont
  {Hall}}, \bibinfo {author} {\bibfnamefont {C.}~\bibnamefont {Hill}}, \bibinfo
  {author} {\bibfnamefont {J.}~\bibnamefont {Cole}}, \bibinfo {author}
  {\bibfnamefont {B.}~\bibnamefont {Städler}}, \bibinfo {author} {\bibfnamefont
  {F.}~\bibnamefont {Caruso}}, \bibinfo {author} {\bibfnamefont
  {P.}~\bibnamefont {Mulvaney}}, \bibinfo {author} {\bibfnamefont
  {J.}~\bibnamefont {Wrachtrup}}, \ and\ \bibinfo {author} {\bibfnamefont
  {L.}~\bibnamefont {Hollenberg}},\ }\href@noop {} {\bibfield  {journal}
  {\bibinfo  {journal} {Proceedings of the National Academy of Sciences}\
  }\textbf {\bibinfo {volume} {107}},\ \bibinfo {pages} {18777} (\bibinfo
  {year} {2010})}\BibitemShut {NoStop}%
\bibitem [{\citenamefont {Hall}\ \emph {et~al.}(2012)\citenamefont {Hall},
  \citenamefont {Beart}, \citenamefont {Thomas}, \citenamefont {Simpson},
  \citenamefont {McGuinness}, \citenamefont {Cole}, \citenamefont {Manton},
  \citenamefont {Scholten}, \citenamefont {Jelezko}, \citenamefont {Wrachtrup}
  \emph {et~al.}}]{hall2012high}%
  \BibitemOpen
  \bibfield  {author} {\bibinfo {author} {\bibfnamefont {L.}~\bibnamefont
  {Hall}}, \bibinfo {author} {\bibfnamefont {G.}~\bibnamefont {Beart}},
  \bibinfo {author} {\bibfnamefont {E.}~\bibnamefont {Thomas}}, \bibinfo
  {author} {\bibfnamefont {D.}~\bibnamefont {Simpson}}, \bibinfo {author}
  {\bibfnamefont {L.}~\bibnamefont {McGuinness}}, \bibinfo {author}
  {\bibfnamefont {J.}~\bibnamefont {Cole}}, \bibinfo {author} {\bibfnamefont
  {J.}~\bibnamefont {Manton}}, \bibinfo {author} {\bibfnamefont
  {R.}~\bibnamefont {Scholten}}, \bibinfo {author} {\bibfnamefont
  {F.}~\bibnamefont {Jelezko}}, \bibinfo {author} {\bibfnamefont
  {J.}~\bibnamefont {Wrachtrup}},  \emph {et~al.},\ }\href@noop {} {\bibfield
  {journal} {\bibinfo  {journal} {Scientific reports}\ }\textbf {\bibinfo
  {volume} {2}} (\bibinfo {year} {2012})}\BibitemShut {NoStop}%
\bibitem [{\citenamefont {Pham}\ \emph {et~al.}(2011)\citenamefont {Pham},
  \citenamefont {{Le Sage}}, \citenamefont {Stanwix}, \citenamefont {Yeung},
  \citenamefont {Glenn}, \citenamefont {Trifonov}, \citenamefont {Cappellaro},
  \citenamefont {Hemmer}, \citenamefont {Lukin}, \citenamefont {Park} \emph
  {et~al.}}]{pham2011magnetic}%
  \BibitemOpen
  \bibfield  {author} {\bibinfo {author} {\bibfnamefont {L.}~\bibnamefont
  {Pham}}, \bibinfo {author} {\bibfnamefont {D.}~\bibnamefont {{Le Sage}}},
  \bibinfo {author} {\bibfnamefont {P.}~\bibnamefont {Stanwix}}, \bibinfo
  {author} {\bibfnamefont {T.}~\bibnamefont {Yeung}}, \bibinfo {author}
  {\bibfnamefont {D.}~\bibnamefont {Glenn}}, \bibinfo {author} {\bibfnamefont
  {A.}~\bibnamefont {Trifonov}}, \bibinfo {author} {\bibfnamefont
  {P.}~\bibnamefont {Cappellaro}}, \bibinfo {author} {\bibfnamefont
  {P.}~\bibnamefont {Hemmer}}, \bibinfo {author} {\bibfnamefont
  {M.}~\bibnamefont {Lukin}}, \bibinfo {author} {\bibfnamefont
  {H.}~\bibnamefont {Park}},  \emph {et~al.},\ }\href@noop {} {\bibfield
  {journal} {\bibinfo  {journal} {New Journal of Physics}\ }\textbf {\bibinfo
  {volume} {13}},\ \bibinfo {pages} {045021} (\bibinfo {year}
  {2011})}\BibitemShut {NoStop}%
\bibitem [{\citenamefont {Kaufmann}\ \emph {et~al.}(2013)\citenamefont
  {Kaufmann}, \citenamefont {Simpson}, \citenamefont {Hall}, \citenamefont
  {Perunicic}, \citenamefont {Senn}, \citenamefont {Steinert}, \citenamefont
  {McGuinness}, \citenamefont {Johnson}, \citenamefont {Ohshima}, \citenamefont
  {Caruso} \emph {et~al.}}]{kaufmann2013detection}%
  \BibitemOpen
  \bibfield  {author} {\bibinfo {author} {\bibfnamefont {S.}~\bibnamefont
  {Kaufmann}}, \bibinfo {author} {\bibfnamefont {D.~A.}\ \bibnamefont
  {Simpson}}, \bibinfo {author} {\bibfnamefont {L.~T.}\ \bibnamefont {Hall}},
  \bibinfo {author} {\bibfnamefont {V.}~\bibnamefont {Perunicic}}, \bibinfo
  {author} {\bibfnamefont {P.}~\bibnamefont {Senn}}, \bibinfo {author}
  {\bibfnamefont {S.}~\bibnamefont {Steinert}}, \bibinfo {author}
  {\bibfnamefont {L.~P.}\ \bibnamefont {McGuinness}}, \bibinfo {author}
  {\bibfnamefont {B.~C.}\ \bibnamefont {Johnson}}, \bibinfo {author}
  {\bibfnamefont {T.}~\bibnamefont {Ohshima}}, \bibinfo {author} {\bibfnamefont
  {F.}~\bibnamefont {Caruso}},  \emph {et~al.},\ }\href@noop {} {\bibfield
  {journal} {\bibinfo  {journal} {arXiv preprint arXiv:1304.3789}\ } (\bibinfo
  {year} {2013})}\BibitemShut {NoStop}%
\bibitem [{\citenamefont {Neumann}\ \emph {et~al.}(2013)\citenamefont
  {Neumann}, \citenamefont {Jakobi}, \citenamefont {Dolde}, \citenamefont
  {Burk}, \citenamefont {Reuter}, \citenamefont {Waldherr}, \citenamefont
  {Honert}, \citenamefont {Wolf}, \citenamefont {Brunner}, \citenamefont {Shim}
  \emph {et~al.}}]{neumann2013high}%
  \BibitemOpen
  \bibfield  {author} {\bibinfo {author} {\bibfnamefont {P.}~\bibnamefont
  {Neumann}}, \bibinfo {author} {\bibfnamefont {I.}~\bibnamefont {Jakobi}},
  \bibinfo {author} {\bibfnamefont {F.}~\bibnamefont {Dolde}}, \bibinfo
  {author} {\bibfnamefont {C.}~\bibnamefont {Burk}}, \bibinfo {author}
  {\bibfnamefont {R.}~\bibnamefont {Reuter}}, \bibinfo {author} {\bibfnamefont
  {G.}~\bibnamefont {Waldherr}}, \bibinfo {author} {\bibfnamefont
  {J.}~\bibnamefont {Honert}}, \bibinfo {author} {\bibfnamefont
  {T.}~\bibnamefont {Wolf}}, \bibinfo {author} {\bibfnamefont {A.}~\bibnamefont
  {Brunner}}, \bibinfo {author} {\bibfnamefont {J.~H.}\ \bibnamefont {Shim}},
  \emph {et~al.},\ }\href@noop {} {\bibfield  {journal} {\bibinfo  {journal}
  {Nano letters}\ } (\bibinfo {year} {2013})}\BibitemShut {NoStop}%
\bibitem [{\citenamefont {Toyli}\ \emph {et~al.}(2013)\citenamefont {Toyli},
  \citenamefont {Charles}, \citenamefont {Christle}, \citenamefont
  {Dobrovitski},\ and\ \citenamefont {Awschalom}}]{toyli2013fluorescence}%
  \BibitemOpen
  \bibfield  {author} {\bibinfo {author} {\bibfnamefont {D.~M.}\ \bibnamefont
  {Toyli}}, \bibinfo {author} {\bibfnamefont {F.}~\bibnamefont {Charles}},
  \bibinfo {author} {\bibfnamefont {D.~J.}\ \bibnamefont {Christle}}, \bibinfo
  {author} {\bibfnamefont {V.~V.}\ \bibnamefont {Dobrovitski}}, \ and\ \bibinfo
  {author} {\bibfnamefont {D.~D.}\ \bibnamefont {Awschalom}},\ }\href@noop {}
  {\bibfield  {journal} {\bibinfo  {journal} {Proceedings of the National
  Academy of Sciences}\ }\textbf {\bibinfo {volume} {110}},\ \bibinfo {pages}
  {8417} (\bibinfo {year} {2013})}\BibitemShut {NoStop}%
\bibitem [{\citenamefont {Kucsko}\ \emph {et~al.}(2013)\citenamefont {Kucsko},
  \citenamefont {Maurer}, \citenamefont {Yao}, \citenamefont {Kubo},
  \citenamefont {Noh}, \citenamefont {Lo}, \citenamefont {Park},\ and\
  \citenamefont {Lukin}}]{kucsko2013nanometre}%
  \BibitemOpen
  \bibfield  {author} {\bibinfo {author} {\bibfnamefont {G.}~\bibnamefont
  {Kucsko}}, \bibinfo {author} {\bibfnamefont {P.}~\bibnamefont {Maurer}},
  \bibinfo {author} {\bibfnamefont {N.}~\bibnamefont {Yao}}, \bibinfo {author}
  {\bibfnamefont {M.}~\bibnamefont {Kubo}}, \bibinfo {author} {\bibfnamefont
  {H.}~\bibnamefont {Noh}}, \bibinfo {author} {\bibfnamefont {P.}~\bibnamefont
  {Lo}}, \bibinfo {author} {\bibfnamefont {H.}~\bibnamefont {Park}}, \ and\
  \bibinfo {author} {\bibfnamefont {M.}~\bibnamefont {Lukin}},\ }\href@noop {}
  {\bibfield  {journal} {\bibinfo  {journal} {Nature}\ }\textbf {\bibinfo
  {volume} {500}},\ \bibinfo {pages} {54} (\bibinfo {year} {2013})}\BibitemShut
  {NoStop}%
\bibitem [{\citenamefont {Doherty}\ \emph {et~al.}(2013)\citenamefont
  {Doherty}, \citenamefont {Manson}, \citenamefont {Delaney}, \citenamefont
  {Jelezko}, \citenamefont {Wrachtrup},\ and\ \citenamefont
  {Hollenberg}}]{doherty2013nitrogen}%
  \BibitemOpen
  \bibfield  {author} {\bibinfo {author} {\bibfnamefont {M.~W.}\ \bibnamefont
  {Doherty}}, \bibinfo {author} {\bibfnamefont {N.~B.}\ \bibnamefont {Manson}},
  \bibinfo {author} {\bibfnamefont {P.}~\bibnamefont {Delaney}}, \bibinfo
  {author} {\bibfnamefont {F.}~\bibnamefont {Jelezko}}, \bibinfo {author}
  {\bibfnamefont {J.}~\bibnamefont {Wrachtrup}}, \ and\ \bibinfo {author}
  {\bibfnamefont {L.~C.}\ \bibnamefont {Hollenberg}},\ }\href@noop {}
  {\bibfield  {journal} {\bibinfo  {journal} {Physics Reports}\ } (\bibinfo
  {year} {2013})}\BibitemShut {NoStop}%
\bibitem [{\citenamefont {Jelezko}\ \emph {et~al.}(2004)\citenamefont
  {Jelezko}, \citenamefont {Gaebel}, \citenamefont {Popa}, \citenamefont
  {Gruber},\ and\ \citenamefont {Wrachtrup}}]{jelezko2004observation}%
  \BibitemOpen
  \bibfield  {author} {\bibinfo {author} {\bibfnamefont {F.}~\bibnamefont
  {Jelezko}}, \bibinfo {author} {\bibfnamefont {T.}~\bibnamefont {Gaebel}},
  \bibinfo {author} {\bibfnamefont {I.}~\bibnamefont {Popa}}, \bibinfo {author}
  {\bibfnamefont {A.}~\bibnamefont {Gruber}}, \ and\ \bibinfo {author}
  {\bibfnamefont {J.}~\bibnamefont {Wrachtrup}},\ }\href@noop {} {\bibfield
  {journal} {\bibinfo  {journal} {Physical Review Letters}\ }\textbf {\bibinfo
  {volume} {92}},\ \bibinfo {pages} {76401} (\bibinfo {year}
  {2004})}\BibitemShut {NoStop}%
\bibitem [{\citenamefont {Balasubramanian}\ \emph {et~al.}(2009)\citenamefont
  {Balasubramanian}, \citenamefont {Neumann}, \citenamefont {Twitchen},
  \citenamefont {Markham}, \citenamefont {Kolesov}, \citenamefont {Mizuochi},
  \citenamefont {Isoya}, \citenamefont {Achard}, \citenamefont {Beck},
  \citenamefont {Tissler} \emph {et~al.}}]{balasubramanian2009ultralong}%
  \BibitemOpen
  \bibfield  {author} {\bibinfo {author} {\bibfnamefont {G.}~\bibnamefont
  {Balasubramanian}}, \bibinfo {author} {\bibfnamefont {P.}~\bibnamefont
  {Neumann}}, \bibinfo {author} {\bibfnamefont {D.}~\bibnamefont {Twitchen}},
  \bibinfo {author} {\bibfnamefont {M.}~\bibnamefont {Markham}}, \bibinfo
  {author} {\bibfnamefont {R.}~\bibnamefont {Kolesov}}, \bibinfo {author}
  {\bibfnamefont {N.}~\bibnamefont {Mizuochi}}, \bibinfo {author}
  {\bibfnamefont {J.}~\bibnamefont {Isoya}}, \bibinfo {author} {\bibfnamefont
  {J.}~\bibnamefont {Achard}}, \bibinfo {author} {\bibfnamefont
  {J.}~\bibnamefont {Beck}}, \bibinfo {author} {\bibfnamefont {J.}~\bibnamefont
  {Tissler}},  \emph {et~al.},\ }\href@noop {} {\bibfield  {journal} {\bibinfo
  {journal} {Nature Materials}\ }\textbf {\bibinfo {volume} {8}},\ \bibinfo
  {pages} {383} (\bibinfo {year} {2009})}\BibitemShut {NoStop}%
\bibitem [{\citenamefont {Maclaurin}\ \emph {et~al.}(2012)\citenamefont
  {Maclaurin}, \citenamefont {Doherty}, \citenamefont {Hollenberg},\ and\
  \citenamefont {Martin}}]{maclaurin2012measurable}%
  \BibitemOpen
  \bibfield  {author} {\bibinfo {author} {\bibfnamefont {D.}~\bibnamefont
  {Maclaurin}}, \bibinfo {author} {\bibfnamefont {M.}~\bibnamefont {Doherty}},
  \bibinfo {author} {\bibfnamefont {L.}~\bibnamefont {Hollenberg}}, \ and\
  \bibinfo {author} {\bibfnamefont {A.}~\bibnamefont {Martin}},\ }\href@noop {}
  {\bibfield  {journal} {\bibinfo  {journal} {Physical Review Letters}\
  }\textbf {\bibinfo {volume} {108}},\ \bibinfo {pages} {240403} (\bibinfo
  {year} {2012})}\BibitemShut {NoStop}%
\bibitem [{\citenamefont {Ledbetter}\ \emph {et~al.}(2012)\citenamefont
  {Ledbetter}, \citenamefont {Jensen}, \citenamefont {Fischer}, \citenamefont
  {Jarmola},\ and\ \citenamefont {Budker}}]{ledbetter2012gyroscopes2}%
  \BibitemOpen
  \bibfield  {author} {\bibinfo {author} {\bibfnamefont {M.}~\bibnamefont
  {Ledbetter}}, \bibinfo {author} {\bibfnamefont {K.}~\bibnamefont {Jensen}},
  \bibinfo {author} {\bibfnamefont {R.}~\bibnamefont {Fischer}}, \bibinfo
  {author} {\bibfnamefont {A.}~\bibnamefont {Jarmola}}, \ and\ \bibinfo
  {author} {\bibfnamefont {D.}~\bibnamefont {Budker}},\ }\href@noop {}
  {\bibfield  {journal} {\bibinfo  {journal} {Physical Review A}\ }\textbf
  {\bibinfo {volume} {86}},\ \bibinfo {pages} {052116} (\bibinfo {year}
  {2012})}\BibitemShut {NoStop}%
\bibitem [{\citenamefont {Maclaurin}\ \emph {et~al.}(2013)\citenamefont
  {Maclaurin}, \citenamefont {Hall}, \citenamefont {Martin},\ and\
  \citenamefont {Hollenberg}}]{maclaurin2013nanoscale}%
  \BibitemOpen
  \bibfield  {author} {\bibinfo {author} {\bibfnamefont {D.}~\bibnamefont
  {Maclaurin}}, \bibinfo {author} {\bibfnamefont {L.}~\bibnamefont {Hall}},
  \bibinfo {author} {\bibfnamefont {A.}~\bibnamefont {Martin}}, \ and\ \bibinfo
  {author} {\bibfnamefont {L.}~\bibnamefont {Hollenberg}},\ }\href@noop {}
  {\bibfield  {journal} {\bibinfo  {journal} {New Journal of Physics}\ }\textbf
  {\bibinfo {volume} {15}},\ \bibinfo {pages} {013041} (\bibinfo {year}
  {2013})}\BibitemShut {NoStop}%
\bibitem [{\citenamefont {{De Chiara}}\ and\ \citenamefont
  {Palma}(2003)}]{de2003berry}%
  \BibitemOpen
  \bibfield  {author} {\bibinfo {author} {\bibfnamefont {G.}~\bibnamefont {{De
  Chiara}}}\ and\ \bibinfo {author} {\bibfnamefont {G.}~\bibnamefont {Palma}},\
  }\href@noop {} {\bibfield  {journal} {\bibinfo  {journal} {Physical Review
  Letters}\ }\textbf {\bibinfo {volume} {91}},\ \bibinfo {pages} {90404}
  (\bibinfo {year} {2003})}\BibitemShut {NoStop}%
\bibitem [{\citenamefont {Clarke}(1982)}]{clarke1982triplet}%
  \BibitemOpen
  \bibfield  {author} {\bibinfo {author} {\bibfnamefont {R.~H.}\ \bibnamefont
  {Clarke}},\ }\href@noop {} {\emph {\bibinfo {title} {{Triplet state ODMR
  spectroscopy: Techniques and applications to biophysical systems}}}}\
  (\bibinfo  {publisher} {Wiley New York},\ \bibinfo {year} {1982})\BibitemShut
  {NoStop}%
\end{thebibliography}%

\end{document}